\documentclass[fleqn,12pt,twoside]{article}
\usepackage{espcrc1}

\usepackage{graphicx}
\usepackage[figuresright]{rotating}

\def\sax{\mbox{SAX J1808.4-3658}}

\newcommand{\AmS}{{\protect\the\textfont2
  A\kern-.1667em\lower.5ex\hbox{M}\kern-.125emS}}

\title{Strangeness in Neutron Stars} 

\author{Ignazio Bombaci\address{Dipartimento di Fisica ``Enrico Fermi'', 
Universit\`a degli Studi di Pisa, \\  
and INFN sezione di Pisa, via Buonarroti, 2,  I-56127, Pisa, Italy }}

\begin{document}

\maketitle

\begin{abstract}
We discuss the role of strangeness on the internal constitution 
and structural properties of neutron stars.   
In particular, we report on recent calculations of  hyperon star properties 
derived from microscopic equations of state for hyperonic matter. 
Next, we discuss the possibility of having a strange quark matter core in 
a neutron star, or the possible existence of strange quark matter stars, 
the so-called strange stars.  
\end{abstract}

\section{INTRODUCTION}
The true nature and the internal constitutions of the ultra-dense  
compact stars known as neutron stars is one of the most fascinating 
enigma in modern astrophysics \cite{gle96,web99,LNP01}.   
Different models for the equation of state (EOS) of dense matter predict 
the neutron star maximum mass $M_{max}$ to be in the range 1.4 -- 2.2~$M_\odot$  
($M_\odot \simeq 2 \times 10^{33}$g is the mass of the Sun), 
and a corresponding central density in the range of  4 -- 8 times  
the normal saturation density ($\rho_0 \sim 2.8 \times 10^{14}$g/cm$^3$) 
of nuclear matter.   
Thus neutron stars are the most likely sites in the universe in which 
strangeness-bearing matter with a strangeness to baryon ratio 
$f_S = -S/B \sim 1$ may exist.  
Strangeness in neutron stars is expected both in a {\it confined}  form 
(hyperons and or kaons), or in a {\it deconfined}  form (strange quark matter).  
Accordingly different types  of ``neutron stars'' are expected theoretically, 
as schematically summariezed in Fig.~1.   

\begin{figure}[htb]
\centerline{
\includegraphics[angle=-90,scale=0.48]{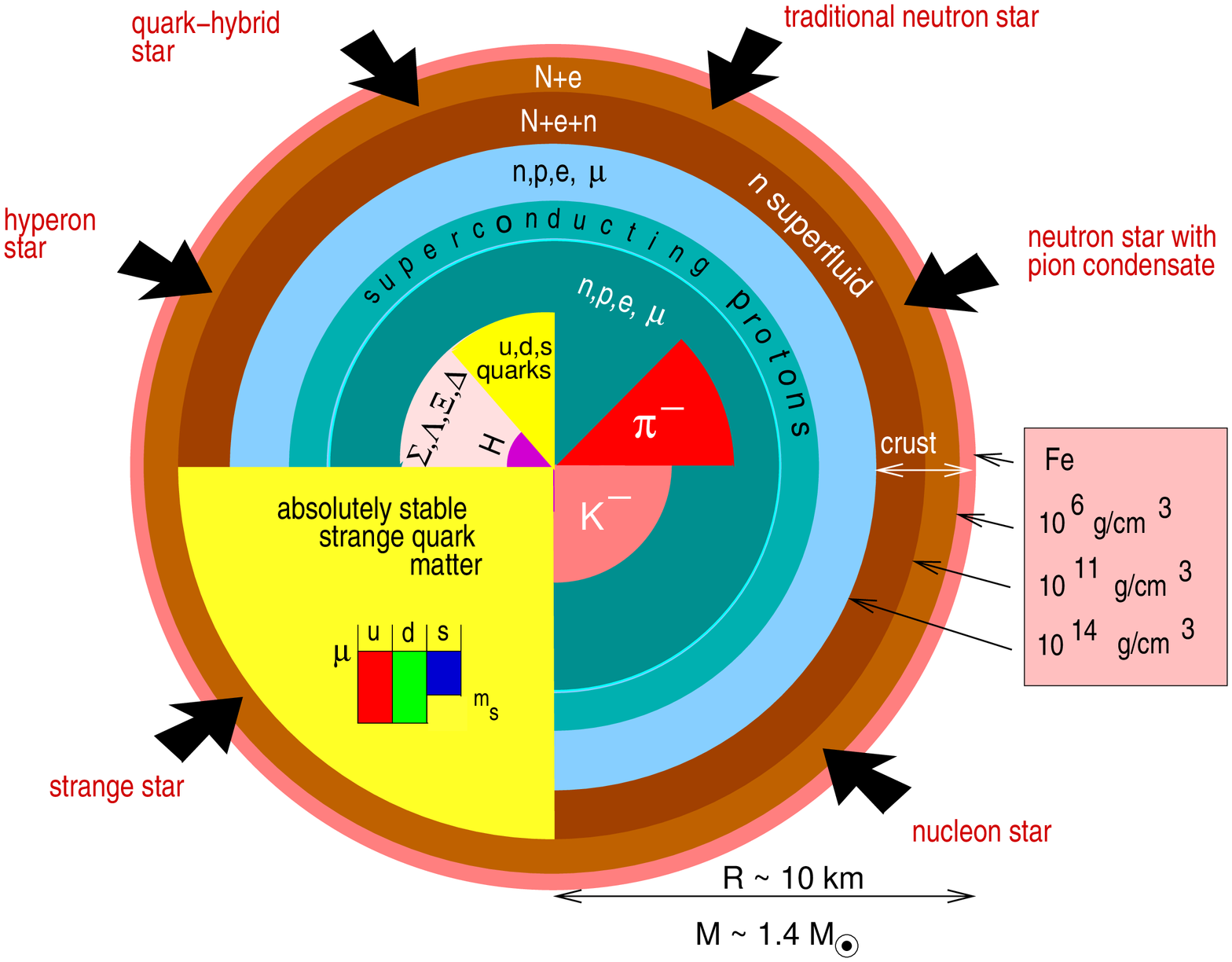}  }
\caption{Schematic cross section of a neutron star according to different 
possibilities for the stellar constituents (adapted from ref. \cite{web99}).}
\label{fig:NStypes}
\end{figure} 

\section{NEUTRON STARS OR HYPERON STARS?}

In a conservative and oversimplified picture the core of a neutron star 
is modeled as a uniform fluid of neutron rich nuclear matter 
in equilibrium with respect to the weak interactions ($\beta$--stable nuclear matter). 
The presence of hyperons in neutron stars was first proposed in 1960 
by Ambartsumyan and Saakyan \cite{AS60}, and since then it has been 
investigated by many authors. 
The reason why hyperons are expected in the high dense core of a neutron star 
is very simple, and it is mainly a consequence of the fermionic nature of nucleons, 
which makes  the nucleon chemical potentials a very  rapidly increasing function 
of density.  As soon as the chemical potential of neutrons becomes sufficiently large 
(see Fig. 2), the most energetic neutrons ({\it i.e.} those on the Fermi surface) 
can decay via the weak interactions into $\Lambda$ hyperons and form a Fermi sea 
of this new hadronic species with $\mu_\Lambda = \mu_n$. 
The $\Sigma^-$ can be produced via the weak process 
$e^- + n \rightarrow \Sigma^- + \nu_e$ when the $\Sigma^-$ chemical potential 
fulfill the  condition\footnote{except from the very initial stage soon after 
neutron star birth, neutrinos freely escape the star and thus the neutrino 
chemical potentials have not to be considered in the chemical equilibrium equations.} 
$\mu_{\Sigma^-} = \mu_n + \mu_e$.   
As we can see from the results depicted in Fig. 2, 
hyperons appear at a relatively moderate density of about 2 times 
the normal saturation density ($n_0 = 0.16$ fm$^{-3}$) of nuclear matter.  
Notice that  the $\Sigma^-$ hyperon appears at a lower density than 
the $\Lambda$, even though the $\Sigma^-$ is more massive than the $\Lambda$.   
This is due to the contribution of the electron chemical potential $\mu_e$ to the 
threshold condition for the $\Sigma^-$  ({\it i.e.} $M_{\Sigma^-} = \mu_n + \mu_e$, 
for free hyperons) and to the fact that $\mu_e$ in dense matter is large 
and can compensate for the  mass difference $M_{\Sigma^-}  -  M_{\Lambda}  = 81.76$ MeV.  

In Fig.~\ref{hypStar}, we show  the profile of a such an hyperon star \cite{Is3}.  
As we see the hyperonic matter inner core of the star extend for about 8~km.   
This radius has to be compared with the total stellar radius $R \sim 11$~km, 
and with the thickness of the nuclear matter layer (outer core)  which is about 2~km.     
Thus neutron stars are ``giant hypernuclei'' \cite{gle85}  under the influence 
of gravity and  strong interactions.     

\begin{figure}[htb]
\centerline{\includegraphics[scale=0.4]{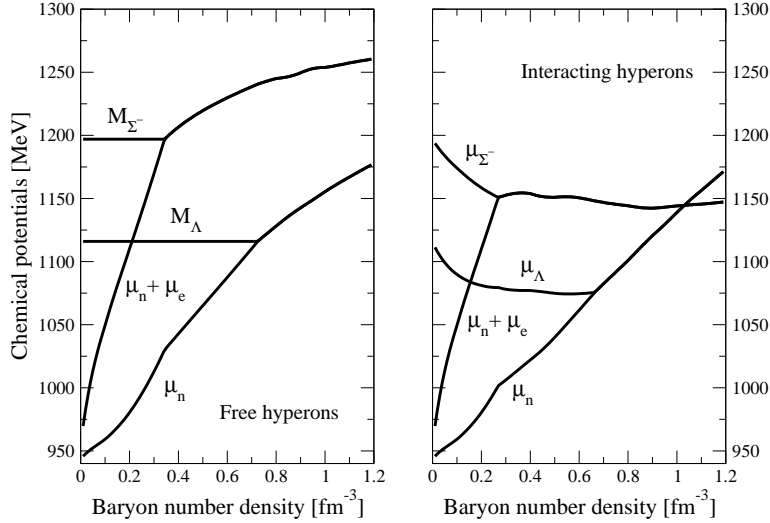}}
\caption{Chemical potentials in $\beta$-stable hyperonic matter. 
Left panel: free hyperons. Right panel:  interacting hyperons (NSC97e interaction). 
Adaped from ref. \cite{vidana}.  
\label{chempot}}
\end{figure} 

\begin{figure}[htb]
\centerline{\includegraphics[scale=0.4]{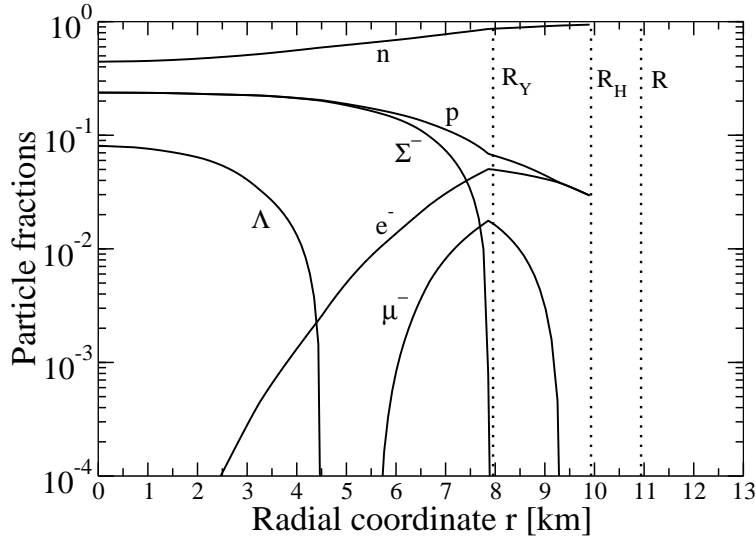}}
\caption{The internal composition of a neutron star with hyperonic matter core. 
The stellar baryonic mass is $M_B = 1.34~M_\odot$. 
$R_Y$ is the radius of the hyperonic core. The nuclear matter layer extend 
between $R_Y$ and $R_H$ and has a thickness of about 2~ km. 
The stellar crust extend between $R_H$ and $R$ and is about 1~km thick.  
(Adapted from Vida\~na {\it et al.}  \cite{Is3}). 
\label{hypStar}}
\end{figure}

The influence of hyperons on neutron stars properties has been investigated 
using different approaches to determine the  EOS of hyperonic matter. 
One of the most popular approaches,  to solve this problem,  is the relativistic 
mean field  model \cite{gle85,ell+91}.     
Some of the parametrizations of the lagrangian of the theory have tried 
to reconcile measurd values of neutron star masses with the binding energy of 
the $\Lambda$ particle in hypernuclei \cite{GM91,HWW99}.    
A different approach is based on the use of local effective potentials to describe 
the in-medium baryon-baryon (BB) interaction \cite{BG97}.     
This method  mimic and generalize to the case of  hyperonic matter  the one 
based on the Skyrme nuclear interaction in the case of nuclear matter. 
Here we will  report on some recent results based on a third approach,  
which starts from the basic BB interaction and solve the many-body  problen 
to get the EOS for hyperonic matter. This method is based on an extension 
of the Brueckener-Bethe-Goldstone (BBG) theory to include hyperonic degrees of 
freedom \cite{hypCT1,hypCT2,Is1,Is2,Is3}.   
In particular, the study of  ref. \cite{Is3} focus on the properties of 
a newborn neutron star, and explore the consequences of neutrino trapping 
in dense matter on the structural properties and on the early evolution of neutron stars 
\cite{physrep}.  

\begin{figure}[htb]
\centerline{\includegraphics[scale=0.42]{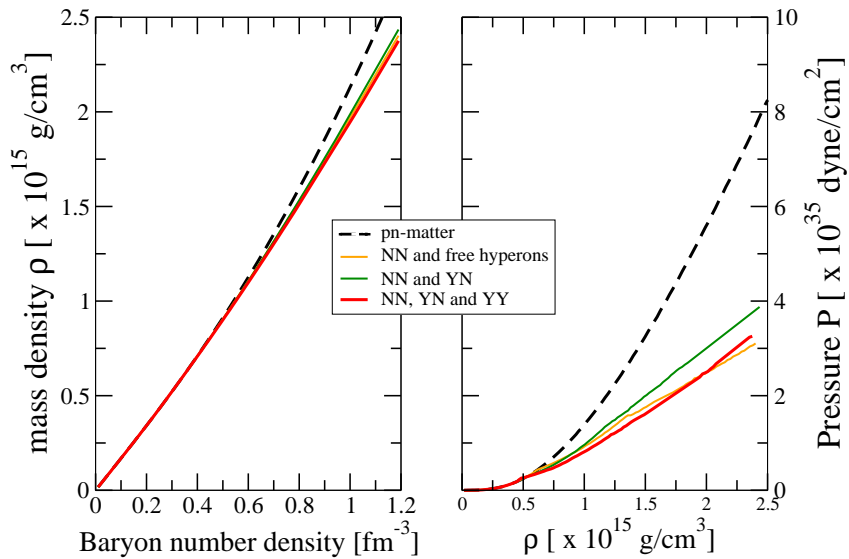}}
\caption{Equation of state of dense hadronic matter with and 
without hyperons \cite{Is2}.}
\label{fig:eos}
\end{figure} 

In  Fig. 4, we show the EOS for $\beta$-stable dense matter 
({\it i.e.} in equilibrium with respect to the weak interactions) 
obtained by Vida\~na {\it et al.} \cite{Is2} using the Nijmegen soft-core potential (NSC97e)  
of ref. \cite{nij97} to describe the hyperon-nucleon (YN) and hyperon-hyperon (YY) 
interaction within the Brueckner-Hartree-Fock (BHF) approximation 
of the extended BBG theory.   
In ref. \cite{Is2}, the pure nucleonic contribution to the EOS has been included 
using  a parametrization of the Akmal-Pandharipande EOS \cite{AP97}, 
where a semi-phenomenological  three-nucleon (NNN) interaction of the Urbana type 
is added to the nuclear hamiltonian to reproduce the empirical saturation point 
of nuclear matter. 

As expected, the presence of hyperons makes the EOS much softer with respect 
to the pure nucleonic case.  
The softening of the EOS caused by the presence of hyperons has important 
consequences on many macroscopic properties of the star: 
the maximum stellar mass is reduced by $\Delta M_{max} \sim$ 0.5 -- 0.8 $M_\odot$, 
and the corresponding central density is increased.  
Also, hyperon stars are more compact ({\it i.e.}  they have a smaller radius) 
with respect to pure nucleonic neutron stars. 
This is illustrated in Fig.~5, where we show the mass radius-relation  
for traditional  neutron stars and for hyperon stars obtained 
with the microscopic EOS of ref. \cite{Is2} (left panel) and with the  
relativistic mean field EOS  (GM3 model) given in ref. \cite{GM91}.   
The results depicted in Fig.~5 clearly demonstrate that to neglect hyperons 
leads to an overstimate of the maximum mass of neutron stars.  

\begin{figure}[htb]
\centerline{\includegraphics[scale=0.45]{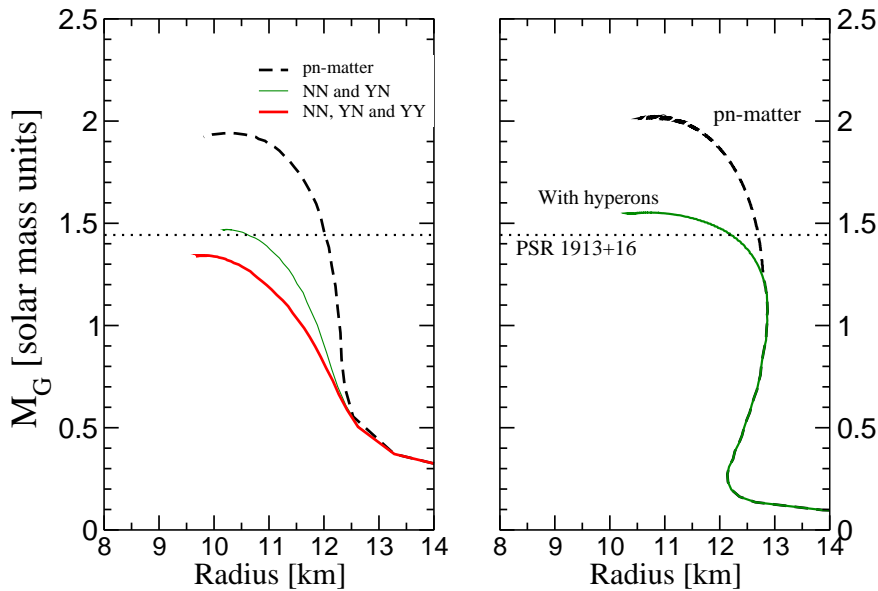}}
\caption{Mass-Radius relation for ``traditional'' neutron stars and hyperon  stars 
calculated \cite{Is2} within the BHF approach with the NSC9e interaction (left panel) 
and  with the relativistic mean field EOS  GM3 of ref. \cite{GM91} (rigth panel).   
The dotted horizontal  line indicates the measured mass for the ``neutron star'' 
in the radio pulsar PSR1913+16 . }
\label{fig:mass-rad}
\end{figure} 

It is important  to notice  the ``low'' value of the stellar maximum mass,  
predicted within the approach of ref. \cite{Is2},    
which is in contrast with some precise determination of neutron star 
masses \cite{kerk}. 
For example in the case of the neutron star associated to the pulsar  
PSR1913+16, the measured stellar mass is \cite{TW89}
\begin{equation} 
      M_{PSR1913+16} = (1.4411 \pm 0.0007)~M_\odot ~.
\label{eq:1913} 
\end{equation} 
The prediction of a value for $M_{max}$ below the measured neutron 
star masses  is a common feature of all the present microscopic EOS of 
hyperonic matter based on G-matrix BHF calculations \cite{hypCT2,Is2,Is3}.  
For example, the authors of ref. \cite{hypCT2}, in 
case of the Argonne $v_{18}$ NN interaction, found 
$M_{max} = 2.00~M_\odot$, a corresponding radius of $R = 10.54$~km and 
a central density $\rho_c = 1.11$~fm$^{-3}$ for neutron stars with a 
pure nucleonic core.  
When hyperons are considered as possible stellar constituents,
 they found \cite{hypCT2} 
$M_{max} = 1.22~M_\odot$, a corresponding radius of $R = 10.46$~km and 
a central density $\rho_c = 1.25$~fm$^{-3}$. 
Therefore the current equations of state  for hyperonic matter, 
deduced from microscopic G-matrix BHF calculations,  
are ``too soft'' to explain observed neutron star masses.   

Clearly, one should try to trace the origin of this problem back to the 
underlying YN and YY two body interactions or to the possible 
repulsive three-body baryonic forces involving one or more hyperons, not 
included in the work of ref.~\cite{hypCT1,hypCT2,Is1,Is2,Is3}. 
Presently this is a subject of very active research by people working in 
this field. Therefore, the use of microscopic EOSs of hyperonic matter in the 
contest of neutron star physics is of fundamental importance for our understanding 
of the strong interactions involving hyperons, and to learn how 
these interactions  behave in dense many-body systems.     
 
\section{KAON CONDENSATION IN NEUTRON STARS} 

The inner core of  neutron stars could also contain a  Bose-Einstein condensate 
of negative kaons \cite{KN86,Brow+94,TPL94,Lee96,GS98}. 
As the density of stellar matter is increased, the $K^-$ energy is lowered by 
the attractive vector mean field originating from dense nucleonic matter. 
When the $K^-$ energy becomes smaller than the electron chemical potential $\mu_e$ 
(which is an increasing function of density) the strangeness changing process 
$ e^- \rightarrow K^- + \nu$~   becomes possible. The critical density 
for this process has been calculated to be 
in the range 2.5 -- 5.0 $n_0$ \cite{TPL94,Lee96}. 

Due to the lack of space, we do no have the possibilty to discuss the 
many relevant implications that kaon condensation  has for the structure 
and the evolution of  neutron stars. 
We refer the reader to the original literature on the subject (see {\it e.g.} 
\cite{KN86,Brow+94,TPL94,Lee96,GS98} and references therein quoted).

\section{HYBRID STARS}  

The core of the more massive neutron stars  is one of the best candidates in 
the Universe where a phase transition from hadronic matter to a deconfined 
quark phase  should  occur. 
The quark-deconfinement phase transition proceeds through a mixed phase 
over a finite range of pressures and densities \cite{gle92,gle96}. 
At the onset of the mixed phase,  quark matter droplets form a Coulomb lattice 
embedded in a sea of hadrons and in a roughly uniform sea of electrons and muons. 
As the pressure increases various geometrical shapes (rods, plates) of the less abundant 
phase immersed in the dominant one are expected. Finally the system turns into 
uniform quark matter  at the highest pressure of the mixed phase \cite{hei93}.  
Compact stars  which possess a quark matter core, either as a 
mixed phase of deconfined quarks and hadrons, or as a pure quark matter phase,  
are called  Hybrid Stars \cite{gle96,web99,LNP01}.     

Many possible astrophysical signals for the appearence of a quark core 
in neutron stars have been proposed in the last few years 
(see \cite{gle96,web99,LNP01} and references therein quoted). 
Particularly,  pulse timing properties of pulsars have attracted much attention 
since they are a manifestation of the rotational properties of the associated 
neutron star. The onset of quark-deconfinement in the core of the star, 
will cause a change in the stellar moment of inertia \cite{GPW97}.  
This change will produce a peculiar evolution of the stellar rotational period  
($P= 2\pi/\Omega$) which will cause large deviations of the so called pulsar 
braking index $n(\Omega) = (\Omega \ddot{\Omega}/\dot{\Omega}^2)$ from the 
{\it canonical} value $n=3$,  derived within the magnetic dipole  model for pulsars 
and assuming a constant moment of inertia for the star.  
The  possible measurement of a value of the braking index very different 
from the canonical value ({\it i.e.} $ |n| >> 3$) has been proposed \cite{GPW97}  
as a signature for the occurrence of the quark-deconfinement phase transition 
in a neutron star.  
However, it must be stressed that a large value of the braking index 
could also results from the pulsar magnetic field decay and/or 
alignment of the magnetic axis with the rotation axis \cite{TK2001}.        

\section{STRANGE STARS} 

The possible existence of a new class of compact stars completely made of 
deconfined {\it u,d,s} quark matter  (strange quark matter (SQM))  is one 
of the consequences of an  hypothesis~\cite{witt} formulated by 
A.R. Bodmer in 1971 and revived by E. Witten in 1984. 
These stars are usually called  Strange Stars. 
According to the Bodmer-Witten hypothesis  SQM  
could be the true ground state of matter.  
In other words, at zero temperature and pressure, the energy per baryon of SQM 
could be less than the energy per baryon of $^{56}$Fe,  which is the most tightly 
bound nucleus in nature. 
The strange matter hypothesis does not conflict with the existence of 
atomic nuclei as conglomerates of nucleons, or with the stability 
of ``ordinary'' matter \cite{fj84,mads99,bomb2001}.  
Thus strange stars may exist in the universe.  

One of the most likely strange star candidate is the compact object in 
the transient X-ray burst source \sax\ (ref.~\cite{Li99a}). 
This X-ray source  was discovered in 1996 by  the  BeppoSAX satellite.   
Two bright type-I X-ray bursts were detected, each lasting less than 
30 seconds.  Analysis of the bursts in \sax\ indicates that it has a peak 
X-ray luminosity of $6\times 10^{36}~$erg/s in its bright state, 
and a X-ray luminosity lower than $10^{35}~$erg/s in quiescence. 
 \sax\ is a X-ray {\it millisecond pulsar} with a pulsation period of 
2.49 ms,  also it is a member of a binary stellar system with 
orbital period of two hours.  
Using the observational data collected by the  Rossi X-ray Timing 
Explorer during the the 1998 April-May outburst,  
Li {\it et al.} \cite{Li99a} have obtained an upper limit for the 
compact star radiud as a function of the unknown stellar mass.   
Comparing this {\it observational} mass-radius  (M-R) relation of \sax\  
with the theoretical M-R realtions for traditional neutron stars, 
hyperon stars, stars with kaon condensation, and strange stars 
Li {\it et al.} \cite{Li99a} (see their Fig. 1) argue 
that a strange star model is more consistent with \sax,  and suggest 
that it could be a strange star.     

\sax\ is  not the only LMXBs which could harbour a strange star. 
Recent studies have shown that the compact stars 
associated with the X-ray burster 4U 1820-30 (ref.\cite{Bomb97}), 
the bursting X-ray pulsar GRO J1744-28 (ref.\cite{Cheng}), the X-ray pulsar 
Her X-1 (ref.\cite{dey98}), the  kHz QPOs source 4U 1728-34 (ref.\cite{Li99b}), 
are likely strange star candidates. 
Recently, it has been suggested that the  isolated compact star  
RX~J1856.5-3754 (ref.~\cite{drake02})  could be a strange star.

\section{QUARK-DECONFINEMENT PHASE TRANSITION IN NEUTRON\\ 
                  STARS AND GAMMA-RAY BURSTS} 

Gamma Ray Bursts (GRBs) are one of the most violent and mysterious 
phenomena in the universe (see {\it e.g.} ref.\cite{piran} for a 
general introduction on this subject).   
During the last ten years two satellites, the Compton Gamma Ray Observatory 
(CGRO) and BeppoSAX, have revolutionized our understanding of GRBs. 
The Burst And Transient Source Experiment (BATSE) on board of the CGRO has 
demonstrated that GRBs originate at cosmological distances. 
The BeppoSAX discovered the X-ray afterglow. 
This has permitted to determine the position of some GRBs,  to identify the 
host galaxy and, in a number of cases, to measure the red-shift. 
If the energy is emitted isotropically, the measured fluence of the bursts  
implies an energy of the order of $10^{53}$ erg.           
However, there is now compelling evidence that the $\gamma$-ray emission is 
not isotropic, but displays a jet-like geometry.  
In this case the GRB energy is of the order of $10^{51}$ erg \cite{fray01}. 

Many cosmological models for the energy source of GRBs have been proposed.  
Presently one of the most popular is  the so-called "collapsar",   
or "hypernova" model.    
Alternative models are the merging of two neutron stars (or a neutron star 
and a black hole) in a binary system, or the accretion of matter 
into a black hole.      
The present report is not the appropriate place to discuss the various 
merits and drawbacks of the many theoretical models for GRBs.  
In the following, we will mention some recent research which try to make a 
connection between GRBs and quark-deconfinement phase transition.  
 
A possible central {\it engine} for GRBs is the conversion of a
pure hadronic compact star  to a strange star. 
The stellar conversion is triggered by the formation of a SQM drop in 
the center of the hadronic star. 
This idea was proposed long time ago by Alcook {\it et al.}  \cite{afo86}.    
Recently detailed calculations, based  on different realistic models for 
the equation of state of neutron star matter and SQM,  
have been performed by the authors of ref.\cite{BD00}.     
They showed that the total amount of energy liberated in the conversion 
is in the range (1 -- 4)$\times 10^{53}$erg. This energy will be 
mainly taken away by the neutrinos produced during the quark-deconfinement 
phase transition. If the efficiency of the conversion of neutrinos to $\gamma$ 
is of the order of a few percent \cite{salm99}, then the birth of a strange star 
from a neutron star could be the energy source for GRBs.  
 
A mounting number of observational data suggest a clear connection between 
supernova (SN) explosions and GRBs \cite{amati,reeves02,bloom99,piro,antonelli}.  
Particularly, in the case of the gamma ray burst of July 5, 1999 
(GRB990705) and in the case of GRB011211, it has been possible to estimate 
the time delay between the two events. 
For GRB990705 the supernova explosion is evaluated to have occurred 
a few years before the GRB \cite{amati,lazzati},  
while for GRB011211 about four days before the burst \cite{reeves02}.  

The scenario which emerges from these findings is the following 
two-stage scenario: 
(i) the first event is the supernova explosion which forms a compact stellar 
remnant, {\it i.e.} a neutron star (NS); 
(ii) the second catastrophic event is associated with the NS and it is the 
energy source for the observed GRB. 
These new observational data, and the scenario outlined above, 
poses severe problems for most of the current theoretical models for 
the central energy source of GRBs. 
The main difficulty of all these models is to give an answer to the 
following questions: what is the origin of the second ``explosion''?    
How to explain the long time delay between the two events?  

In the so-called {\it supranova} model \cite{vietri} for GRBs  
the second catastrophic event is the collapse to a black hole of a 
{\it supramassive} neutron star, {\it i.e.} a fast rotating NS 
with a baryonic mass $M_B$ above the maximum baryonic mass $M_{B,max}$ 
for non-rotating 
configurations. In this model, the time delay between the SN explosion and 
the GRB is equal to the time needed by the fast rotating newly formed 
neutron star to get rid of angular momentum and to reach the limit for 
instability against quasi-radial modes where the collapse to a black hole 
occurs \cite{DTB98}.  
The supranova model needs a fine tuning in the initial spin period $P_{in}$ 
and baryonic stellar mass $M_{B,in}$ to produce a supramassive neutron star 
that can be stabilized by rotation up to a few years. 
For example, if $P_{in} \geq 1.5$~ms, then the newborn supramassive neutron star 
must be formed within $\sim 0.03 M_\odot$  above  $M_{B,max}$  \cite{DTB98}.

In a very recent paper, Berezhiani {\it et al.} \cite{BBDFL02} (see also ref. \cite{iii})     
have proposed a new model to explain the SN--GRB association and in 
particular the long time delay inferred for GRB990705 and GRB011211.   
In the model of ref.\cite{BBDFL02}, the second explosion is related to 
the conversion from a metastable purely Hadronic Star (neutron star or 
hyperon star) into a more compact star in which deconfined quark matter  
is present ({\it i.e.} a hybrid  star or a strange star).    
The new and crucial idea in the work of ref. \cite{BBDFL02} 
with respect to previous work \cite{BD00}, is the metastability  
of the hadronic star due to the existence of a non-vanishing 
surface tension at the interface separating hadronic matter
from quark matter. The {\it mean-life time} of the metastable hadronic star 
can then be connected to the delay between the SN explosion and 
the GRB. 
The nucleation time ({\it i.e.} the time to form a critical-size drop of 
quark matter) can be extremely long if the mass of the star is small.  
Via mass accretion the nucleation time can be dramatically reduced 
and the star is finally converted from the metastable into the stable 
configuration \cite{BBDFL02,iii,isaac}.     
A huge amount of energy, of the order of 10$^{52}$--10$^{53}$ erg, 
is released during the conversion process and can produce 
a powerful gamma ray burst. 
Within the model proposed by Berezhiani {\it et al.} \cite{BBDFL02}
is is possible to have different time delays between the two events 
since the {\it mean-life time} of the metastable hadronic star    
depends  on the value of the stellar central pressure. 
Thus the model of ref. \cite{BBDFL02} is able to interpret 
a time delay of a few years (as observed in GRB990705 \cite{amati,lazzati}),  
of a few days (as in the case of GRB011211 \cite{reeves02}), 
or the nearly simultaneity of the two events (as in the case of 
SN2003dh and GRB030329 \cite{hjorth03}).

\section*{Acknowledgments} 
I am greatly indebted to  I. Vida\~na for providing me some unpublished results 
from his Ph.D. thesis and for the great help in preparing most of the figures 
of the present work.

\end{document}